\documentclass{article}
\usepackage{authblk}
\usepackage{tabularx}
\usepackage[margin=1in]{geometry}
\usepackage[onehalfspacing]{setspace}
\usepackage{amsmath,mathtools}
\usepackage{pgfplots}
\usepackage{subcaption}
\usepackage{booktabs}
\usepackage[braket, qm]{qcircuit}
\usepackage{graphicx}
\usepackage{placeins}
\usepackage{authblk}
 \usepackage{xcolor}
\usepackage{soul}
\usepackage[linesnumbered,ruled,vlined]{algorithm2e}

\SetCommentSty{mycommfont}

\SetKwInput{KwInput}{Input}                
\SetKwInput{KwOutput}{Output}              
\SetKwProg{Fn}{Function}{}{}

\usepackage[english]{babel}

\usepackage{pdflscape}
\usepackage{fancyvrb}
\usepackage{calc}
\usepackage{rotating}
\usepackage{pgf,tikz}
\usepackage{mathrsfs}
\usetikzlibrary{arrows}
\usepackage{mathtools}

\usepackage{listings}
\usepackage[]{qcircuit}
\usepackage{braket}
\usepackage{mathtools}
\usepackage{varwidth}
\usepackage{caption}
\usepackage{subcaption}
\usepackage{graphicx}
\usepackage[export]{adjustbox}

\usepackage{tcolorbox}

\newcommand{\norm}[1]{\left\lVert#1\right\rVert}

\makeatletter
\def\paragraph{\@startsection{paragraph}{4}%
	\z@\z@{-\fontdimen2\font}%
	{\normalfont\bfseries}}
\makeatother

\usepackage{graphicx}
\usepackage{pgffor}
\usepackage{tikz,tikz-cd}
\usetikzlibrary{backgrounds,fit,decorations.pathreplacing}  
\usepackage{booktabs}
\usepackage{enumerate}

\usepackage{amssymb, amsmath, amsthm}

\usepackage{cite}

\usepackage{xypic}
\usepackage{scalerel}
\usepackage{ulem}

\usepackage{graphicx}              
\usepackage{amsmath}               
\usepackage{amsfonts}              
\usepackage{amsthm}                
\usepackage{xkeyval}               
\usepackage{amssymb}
\usepackage{enumerate}             
\usepackage{color}
\usepackage{breqn}                 
\usepackage{bm}                    

\allowdisplaybreaks[1]
\usepackage{nicefrac}
\usepackage{booktabs}

\usepackage{etoolbox}

\usepackage{kpfonts}
\usepackage{stackengine}
\usepackage{calc}
\newlength\shlength
\newcommand\xshlongvec[2][0]{\setlength\shlength{#1pt}%
	\stackengine{-5.6pt}{$#2$}{\smash{$\kern\shlength%
			\stackengine{7.55pt}{$\mathchar"017E$}%
			{\rule{\widthof{$#2$}}{.57pt}\kern.4pt}{O}{r}{F}{F}{L}\kern-\shlength$}}%
	{O}{c}{F}{T}{S}}

\usepackage{hyperref}
\hypersetup{
	colorlinks   = true,    
	citecolor    = red      
}

\newcommand{\RN}[1]{%
	\textup{\uppercase\expandafter{\romannumeral#1}}%
}

\allowdisplaybreaks

\newcommand{\meqref}[1]{\text{Eq}.~\eqref{#1}}
\newcommand{\mref}[1]{Sec.~$ \!\ref{#1} $}
\newcommand{\mfig}[1]{Fig.~$ \!\ref{#1} $}

\newtheorem{thm}{Theorem}[subsection]

\newtheorem{remark}[thm]{Remark}

\arraycolsep=3pt\def\arraystretch{0.5}

\DeclareMathOperator{\Sfun}{S}

\def\<{\langle}
\def\>{\rangle}

\usetikzlibrary{arrows.meta, quotes}
\makeatletter
\tikzset{
  node on line/.style={
    to path={
      \pgfextra{%
        \edef\tikz@temp{
          edge[
            line to, path only, 
            every edge quotes/.append style={auto=false},
            nodes={alias=@nodeonline@}]
          coordinate(@nodeonline@)
          \unexpanded\expandafter{\tikz@tonodes}(\tikztotarget)
        }\expandafter
      }\tikz@temp
      -- (@nodeonline@) -- (\tikztotarget)}}}
\makeatother

\numberwithin{equation}{section}

\usepackage{bitset}
\pgfplotsset{compat=1.18}
\usepackage{float}

\begin{document}

 \title{Quantum algorithm for edge detection in digital grayscale images}

	\author{Mohit Rohida}
	\author[2]{Alok Shukla}
	 \author[3]{Prakash Vedula}
	\affil{School of Arts and Sciences, Ahmedabad University, India}
    \affil[3]{School of Aerospace and Mechanical Engineering, University of Oklahoma, USA}
	\affil{mohit.r@ahduni.edu.in}
	\affil[2]{alok.shukla@ahduni.edu.in}
	 \affil[3]{pvedula@ou.edu}
 
        \maketitle

\begin{abstract}
In this work, we propose a novel quantum algorithm for edge detection in digital grayscale images, based on the sequency-ordered Walsh-Hadamard transform. The proposed method significantly improves upon existing quantum techniques for edge detection by using a quantum algorithm for the sequency-ordered Walsh-Hadamard transform, achieving a circuit depth of $\mathcal{O}(n)$ (where $n$ is the number of qubits). This represents a notable enhancement over the Quantum Fourier Transform (QFT), which has a circuit depth of $\mathcal{O}(n^{2})$. Furthermore, our approach for edge detection has a computational cost (both gate complexity and quantum circuit depth) of $\mathcal{O}(\log_{2}(N_{1}N_{2}))$ for an image of size $N_{1}\times N_{2}$, offering a considerable improvement over the Quantum Hadamard Edge Detection (QHED) algorithm, which incurs a cost of $\mathcal{O}(\text{poly}(\log_{2}(N_{1}N_{2})))$. By integrating a quantum high-pass filter with the sequency-ordered Walsh-Hadamard transform, the algorithm effectively extracts edge information from images. Computational examples are provided to demonstrate the efficacy of the proposed algorithm which provides a better performance in comparison to QHED. 
\end{abstract}

\section{Introduction}\label{sec:introduction}
The field of quantum computing has rapidly expanded, demonstrating its potential across diverse scientific and engineering disciplines. Notable advancements include Grover's algorithm for unstructured database search \cite{grover1996fast, shukla2025efficientsearch}, Shor's algorithm for integer factorization \cite{shor1994algorithms, shor1999polynomial}, and the Harrow-Hassidim-Lloyd (HHL) algorithm for solving linear systems of equations \cite{harrow2009quantum}. Beyond these foundational algorithms, quantum approaches are also being explored for solving linear and non-linear ordinary differential equations \cite{shukla2021hybrid, berry2014high, childs2020quantum}, numerical integration \cite{shu2024general, shukla2024partialsum}, image processing \cite{rohida2024hybrid, shukla2022hybrid, yao2017quantum}, and non-linear optimal control problems \cite{sandesara2025quantum}. This work specifically focuses on the application of quantum algorithms to image processing, a domain increasingly challenged by the computational demands of high-resolution images prevalent in astronomy, medicine, and engineering \cite{rohida2024hybrid, shukla2022hybrid, yao2017quantum, wang2022review, ruan2021quantum}. Our particular interest lies in edge detection, a fundamental image processing task that identifies sharp intensity changes between adjacent pixels, crucial for object recognition, segmentation, and feature extraction \cite{gonzalez2009digital}.

Classical edge detection algorithms often rely on transforming images into the Fourier domain to suppress low-frequency components, thereby highlighting high-frequency components that correspond to edges \cite{gonzalez2009digital}. Building upon this concept, our proposed quantum algorithm operates in the Walsh domain. 
Walsh basis functions have been widely applied across various disciplines, including in numerical solution of differential equations \cite{shukla2021hybrid}, data compression, image and signal processing \cite{beauchamp1975Walsh, shukla2022hybrid, shukla2023quantum, rohida2024hybrid}. 
Recent research has extended these basis functions, demonstrating their potential for impactful quantum applications, including digital signal filtering \cite{shukla2025generalizedtt}.
In the Walsh domain, a digital image is characterized by high and low sequency components (\mref{sec:WHT}). Analogous to the frequency domain, high-sequency components in the Walsh domain carry information about image edges. We employ a quantum high-pass filter designed to pass these high-sequency components while suppressing low-sequency ones, resulting in an edge-detected image. A significant advantage of using the Walsh-Hadamard transform is its inherent efficiency: a natural-ordered Walsh-Hadamard transform can be implemented with a quantum circuit depth of $\mathcal{O}(1)$ using Hadamard gates. This contrasts sharply with the Quantum Fourier Transform (QFT), which  requires an $\mathcal{O}(n^2)$ quantum circuit depth, where $n$ is the number of qubits.

For the specific approach presented here, a sequency-ordered Walsh-Hadamard transform is required. To achieve this, we integrate a unitary operator $U_z$ (\mref{sec:WHT}) adapted from \cite{shukla2023quantum}, which converts the natural-ordered Walsh-Hadamard transformed vector into its sequency-ordered counterpart. The output of this sequency-ordered transform is then processed by a quantum high-pass filter circuit to isolate edge information by suppressing low-sequency components. State preparation is a critical initial step in any quantum computation, and various quantum image representations have been proposed \cite{yan2016survey}. For our algorithm, we use Quantum Probability Image Encoding (QPIE), a representation similar to that employed in the Quantum Hadamard Edge Detection (QHED) algorithm \cite{yao2017quantum}. We further validate our proposed edge detection algorithm by providing computational examples and comparing its performance with QHED using the Structural Similarity Index Measure (SSIM) as the evaluation metric.

The structure of the remainder of this article is as follows. Section \ref{sec:state_preparation} discusses the Quantum Probability Image Encoding (QPIE) method for state preparation. Section \ref{sec:WHT} provides an introduction to the Walsh-Hadamard transform and describes the method adopted from \cite{shukla2023quantum} for obtaining the sequency-ordered transform. Section \ref{sec:high_pass_filter} presents the quantum high-pass filter, a key component of our proposed edge detection method, which is elaborated in  \mref{sec:edge_detection_method}. Section \ref{subsec:computational_examples} offers several computational examples and a comparison with QHED. Finally,  \mref{sec:conclusion} summarizes the paper's key findings.

\subsection{Notation}

\begin{itemize}
	\item $ \oplus $ : $ x \oplus y $ denotes $ x + y \pmod 2 $.
	\item $ j \cdot k $ : Let $ j = j_{n-1}\,j_{n-2}\,\ldots \, j_1\, j_0 $ and $ k =  k_{n-1}\,k_{n-2} \,\ldots \, k_1\, k_0 $ be the binary notations for $j$ and $k$ with $ j_i,\, k_i \in \{0,1\}$, $ j \cdot k $ denotes the bit-wise dot product of  $j $ and $ k $ modulo $ 2 $, i.e.,  $j \cdot k :=  j_{0}k_{0} + j_{1}k_{1}+ \ldots + j_{n-1}k_{n-1} \pmod 2 $.
	\end{itemize}

\section{State preparation}\label{sec:state_preparation}

The initial step in any quantum image processing algorithm is to encode classical image data into a quantum state. This process, known as quantum image state preparation, is crucial as it dictates how the image information is represented and subsequently manipulated within the quantum domain. Various quantum image representations have been developed to efficiently store and process image data on quantum computers \cite{yan2016survey}. For our proposed quantum edge detection algorithm, we use the Quantum Probability Image Encoding (QPIE) method, which allows for the representation of an $N_1 \times N_2$ grayscale image as a quantum state. This encoding method is well-suited for our purposes, being similar to the representation employed in the Quantum Hadamard Edge Detection (QHED) algorithm \cite{yao2017quantum}.

Consider a classical grayscale image $I$ with dimensions $N_1 \times N_2$, where each pixel $(x, y)$ has an intensity value $P(x, y)$ ranging from $0$ to $2^k-1$ for a $k$-bit image. In QPIE, this image is normalized such that the sum of squares of all normalized pixel intensities equates to one, effectively treating the pixel values as components of a unit vector. This normalization is essential for preparing a valid quantum state. The normalized pixel intensity at position $(x, y)$ can be denoted as 
\begin{equation}
    \tilde{P}(x, y) = P(x, y) / \Sfun,
\end{equation}
where $\Sfun = \sqrt{\sum_{i=0}^{N_1-1}\sum_{j=0}^{N_2-1} P^2(i, j)}$.
We construct a vector $f$ by ``flattening'' the image by sequentially combining the columns of the normalized image intensity matrix as follows:
\begin{equation}
f =  [\tilde{P}(0,0),\tilde{P}(1,0),\ldots,\tilde{P}(N_1 -1,0),\tilde{P}(0,1),\tilde{P}(1,1),\ldots,\tilde{P}(N_1-1,1),\ldots,\tilde{P}(0,N_2 - 1)\ldots,\tilde{P}(N_1-1,N_2-1)]^T. 
\end{equation}

The elements of $f$ are used to prepare an state vector $\ket{\psi}$, such that 
\begin{equation}\label{eq:QPIE_state}
    \ket{\psi} = \sum_{k=0}^{2^n-1} f_k \ket{k}, \quad \text{where}~n=n_1+n_2=\text{log}_2N_1+\text{log}_2N_2.
\end{equation}

The state $\ket{\psi}$ is the quantum image representation prepared for the input image with QPIE. It is easy to see that the number of qubits required to prepare a quantum image representation for a given image $F$ (of size $2^{n_1} \times 2^{n_2}$) with QPIE is $n=n_1+n_2$. For our proposed edge detection approach in grayscale images, we would require $n+1$ qubits with one qubit acting as an ancilla qubit (refer \mfig{proposed_method_block_diagram}). Further, we use a quantum circuit for the sequency-ordered Walsh-Hadamard transform for the state $\ket{\psi}$, which is discussed in the following section.

\section{Walsh-Hadamard transform for sequency-ordered Walsh basis functions}\label{sec:WHT}

Edge detection in classical image processing often relies on transforming images into the Fourier domain to isolate and enhance high-frequency components, which correspond to sharp intensity changes or edges \cite{gonzalez2009digital}. Similarly, our proposed quantum edge detection algorithm operates within the Walsh domain, obtained via the Walsh-Hadamard transform (WHT). The selection of the Walsh domain is motivated by the fact that quantum algorithms for the WHT exhibit a lower quantum circuit depth ($\mathcal{O}(n)$, where $n$ is the number of qubits) compared to those for the Quantum Fourier Transform (QFT) ($\mathcal{O}(n^2)$).

\begin{figure}[htbp]
  \centering

  \begin{minipage}{0.23\textwidth}
    \centering
    \includegraphics[width=\linewidth]{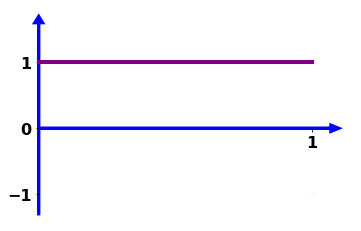}
    \subcaption{$W_0^{(8)}(x)$ }
  \end{minipage}
  \hfill
  \begin{minipage}{0.23\textwidth}
    \centering
    \includegraphics[width=\linewidth]{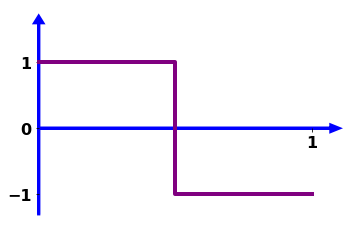}
    \subcaption{$W_1^{(8)}(x)$ }
  \end{minipage}
  \hfill
  \begin{minipage}{0.23\textwidth}
    \centering
    \includegraphics[width=\linewidth]{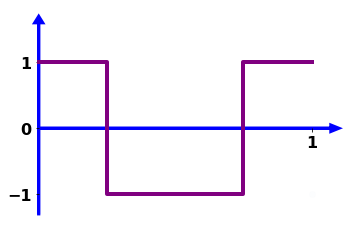}
    \subcaption{$W_2^{(8)}(x)$ }
  \end{minipage}
  \hfill
  \begin{minipage}{0.23\textwidth}
    \centering
    \includegraphics[width=\linewidth]{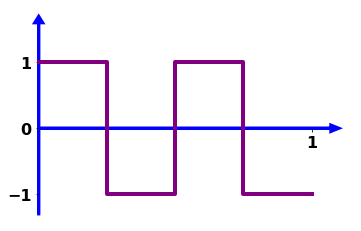}
    \subcaption{$W_3^{(8)}(x)$ }
  \end{minipage}

  \vspace{1em}

  \begin{minipage}{0.23\textwidth}
    \centering
    \includegraphics[width=\linewidth]{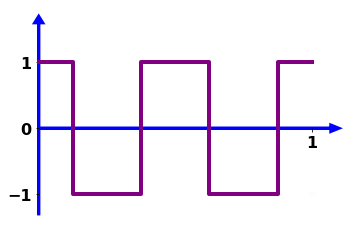}
    \subcaption{$W_4^{(8)}(x)$ }
  \end{minipage}
  \hfill
  \begin{minipage}{0.23\textwidth}
    \centering
    \includegraphics[width=\linewidth]{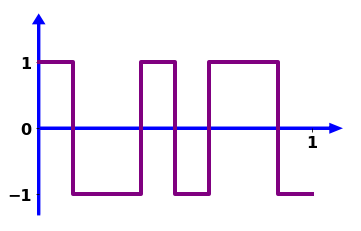}
    \subcaption{$W_5^{(8)}(x)$ }
  \end{minipage}
  \hfill
  \begin{minipage}{0.23\textwidth}
    \centering
    \includegraphics[width=\linewidth]{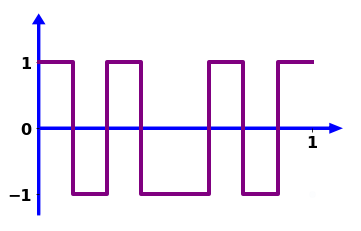}
    \subcaption{$W_6^{(8)}(x)$ }
  \end{minipage}
  \hfill
  \begin{minipage}{0.23\textwidth}
    \centering
    \includegraphics[width=\linewidth]{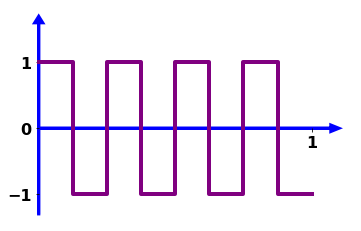}
    \subcaption{$W_7^{(8)}(x)$ }
  \end{minipage}

  \caption{Walsh basis functions in sequency order labeled from $W_0^{(8)}(x)$  to $W_7^{(8)}(x)$. }

\label{wh_basis_sequency_N8}
  
\end{figure}

The WHT decomposes a signal into a set of orthogonal Walsh basis functions. For applications such as spectral analysis and filtering in image processing, a specific ordering of these basis functions, known as sequency ordering, is highly advantageous. Sequency is analogous to frequency in Fourier analysis, representing the number of zero-crossings in a Walsh function \cite{shukla2023quantum, shukla2024sequency, shukla2022quantum}. In this ordering, high-sequency components naturally correspond to fine details and edges within an image. Figure \ref{wh_basis_sequency_N8} shows Walsh basis functions in sequency order for $N=8$.

The sequency-ordered Walsh-Hadamard transform, denoted as $H_N^S:V\rightarrow V$, maps a quantum state in an $N$-dimensional complex vector space $V$ (with computational basis $\{\ket{0},\ket{1},\ldots,\ket{N-1}\}$) to its Walsh domain representation. For an input state $\ket{k}$, where $N = 2^n$ and $n \in \mathbb{N}$ is the number of qubits, the transform is given by:
\begin{equation} \label{eq:sequency_WHT}
H_N^S\ket{k} = \frac{1}{\sqrt{N}}  \sum_{m = 0}^{N -1} \,(-1)^{\sum_{r=0}^{n-1}m_{n-1-r}(k_r \oplus k_{r+1})}\, \ket{m}.
\end{equation}
Here, $k_r$ represents the $r$-th bit of $k$, and $m_{n-1-r}$ represents the $(n-1-r)$-th bit of $m$. The notation $j \oplus k$ represents addition modulo 2 (XOR operation). For an $N=8$ ($n=3$) system, the matrix form of the sequency-ordered Walsh-Hadamard transform is:
\begin{align*}
H^S_8 =
	\frac{1}{\sqrt{8}} \,
	\begin{pmatrix*}[r]
		1 & 1 & 1 & 1 & 1 & 1 & 1 & 1  \\
		1 & 1 & 1 & 1 & -1 & -1 & -1 & -1  \\
		1 & 1 & -1 & -1 & -1 & -1 & 1 & 1  \\
		1 & 1 & -1 & -1 & 1 & 1 & -1 & -1  \\
		1 & -1 & -1 & 1 & 1 & -1 & -1 & 1  \\
		1 & -1 & -1 & 1 & -1 & 1 & 1 & -1  \\
		1 & -1 & 1 & -1 & -1 & 1 & -1 & 1  \\
		1 & -1 & 1 & -1 & 1 & -1 & 1 & -1  \\
	\end{pmatrix*}.
\end{align*}
In contrast, the natural-ordered Walsh-Hadamard transform, $H_N$, is defined by the action:
\begin{equation}
    H_N\ket{k} = \frac{1}{\sqrt{N}}  \sum_{m = 0}^{N -1} \,(-1)^{m\cdot k }\, \ket{m},
\end{equation}
where $m \cdot k$ denotes the bit-wise dot product modulo 2. For $N=8$, the matrix form of the natural-ordered Walsh-Hadamard transform is:
\begin{align*}
H_8 =
	\frac{1}{\sqrt{8}} \,
	\begin{pmatrix*}[r]
		1 & 1  & 1  & 1  & 1  & 1  & 1  & 1 \\
		1 & -1 & 1  & -1 & 1  & -1 & 1  & -1  \\
		1 & 1  & -1 & -1 & 1  & 1  & -1 & -1  \\
		1 & -1 & -1 & 1  & 1  & -1 & -1 & 1  \\
		1 & 1  & 1  & 1  & -1 & -1 & -1 & -1  \\
		1 & -1 & 1  & -1 & -1 & 1  & -1 & 1  \\
		1 & 1  & -1 & -1 & -1 & -1 & 1  & 1  \\
		1 & -1 & -1 & 1  & -1 & 1  & 1  & -1  \\
	\end{pmatrix*}.
\end{align*}

\begin{figure}
    \centering
    \includegraphics[width=0.7\linewidth]{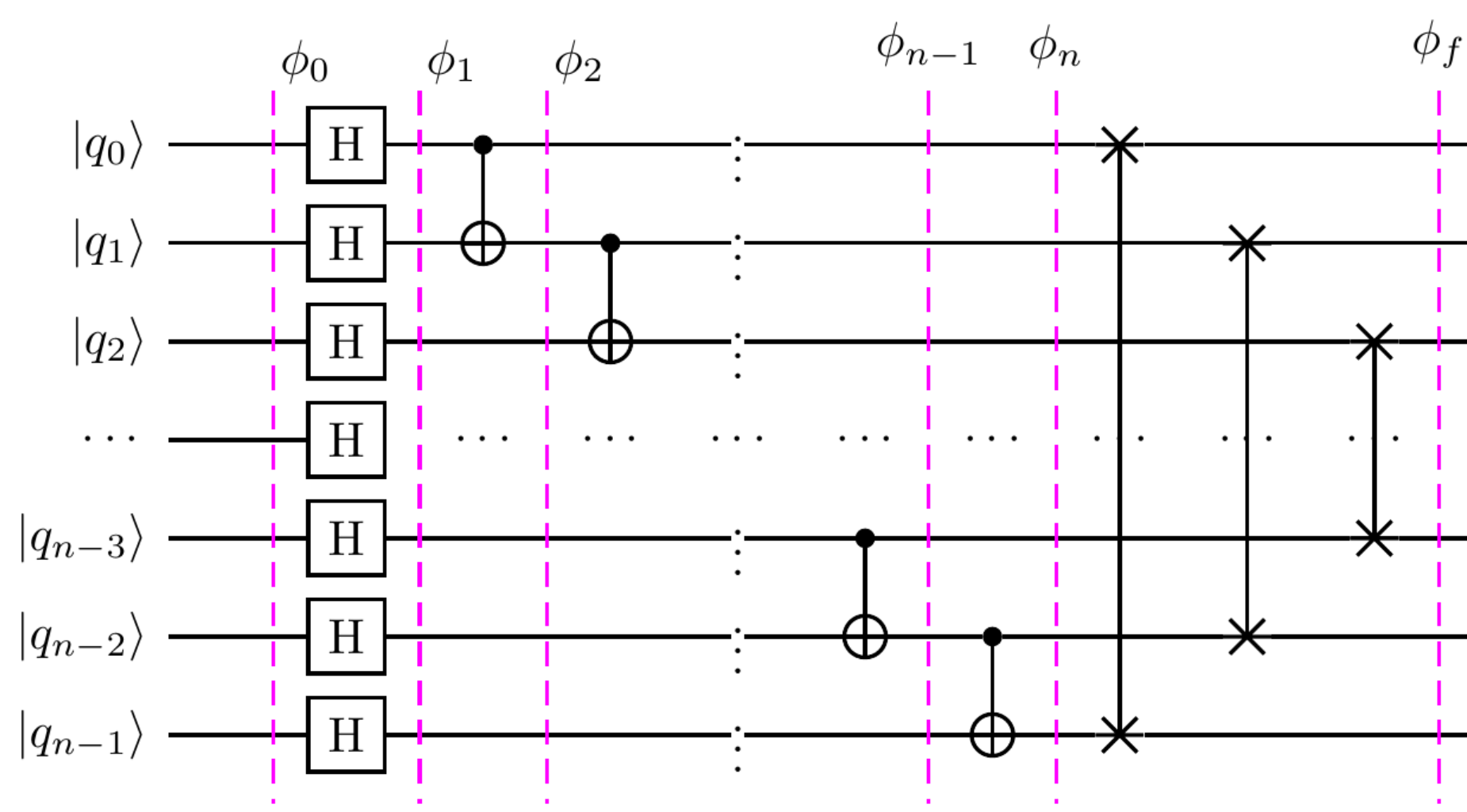}
    \caption{Quantum circuit for sequency-ordered Walsh-Hadamard transform.}\label{fig:sequency_WHT}
\end{figure}

On a quantum computer, the natural-ordered Walsh-Hadamard transform $H_N$ can be efficiently implemented by applying a Hadamard gate to each of the $n$ qubits. For an $n$-qubit input state $\ket{k}$, the transformed state $\ket{\widehat{k}}$ is:
\begin{equation}
  \ket{\widehat{k}} = H^{\otimes n}\ket{k}.
\end{equation}
This operation has a remarkable quantum circuit depth of $\mathcal{O}(1)$, making it exceptionally efficient. To achieve a sequency-ordered Walsh-Hadamard transform, following \cite{shukla2023quantum} we introduce a unitary operator $U_z$, which transforms a natural-ordered Walsh-Hadamard state into its sequency-ordered counterpart. The action of $U_z$ on a computational basis state $\ket{m}$ is defined as:
\begin{equation}
U_z\ket{m}=\ket{g}, \text{~where~} g = \sum_{i=0}^{n-1}2^ig_i \text{~and~} g_i = m_0 \oplus m_1\oplus m_2 \oplus \ldots \oplus m_{n-i-1}.
\end{equation}
It can be shown that $m_i=g_{n-i}\oplus g_{n-i-1}$. Consider an $n$-qubit state $\ket{k}$ for which we seek the sequency-ordered Walsh-Hadamard transform, denoted by $\ket{\widehat{k}}_s$. Applying $U_z$ to the natural-ordered Walsh-Hadamard transformed state $H^{\otimes n}\ket{k}$ yields:
\begin{align}
    \ket{\widehat{k}}_s &= U_z\left(H^{\otimes n}\ket{k}\right) = U_z \left(\frac{1}{\sqrt{N}}  \sum_{m = 0}^{N -1} \,(-1)^{m\cdot k }\, \ket{m}\right) \nonumber\\
    &= \frac{1}{\sqrt{N}}  \sum_{g = 0}^{N -1} (-1)^{\sum_{i=0}^{n-1} k_i(g_{n-i}\oplus g_{n-i-1})} \ket{g} \nonumber\\
    &= \frac{1}{\sqrt{N}}  \sum_{g = 0}^{N -1} (-1)^{\sum_{r=0}^{n-1} k_{n-1-r}(g_{r}\oplus g_{r+1})} \ket{g} \label{eq:proof_2}.
\end{align}
By comparing \meqref{eq:proof_2} with the definition of the sequency-ordered Walsh-Hadamard transform in \meqref{eq:sequency_WHT}, it is evident that $U_z$ is the unitary operator required to convert a natural-ordered Walsh-Hadamard transform to its sequency-ordered form. Thus:
\begin{equation}
    \ket{\widehat{k}}_s = H_N^S\ket{k} = U_zH^{\otimes n}\ket{k} = U_z\ket{\widehat{k}}.
\end{equation}
The complete quantum circuit for obtaining a sequency-ordered Walsh-Hadamard transform is illustrated in \mfig{fig:sequency_WHT}. This circuit achieves a quantum depth of $\mathcal{O}(n)$ for an $n$-qubit system, representing a significant advantage over the $\mathcal{O}(n^2)$ depth typically associated with the Quantum Fourier Transform. It is straightforward to show that the inverse sequency-ordered Walsh-Hadamard transform is performed by the operator $H^{\otimes n}U^{-1}_z$:
\begin{equation}
    \ket{k} = H^{\otimes n}U^{-1}_z\ket{\widehat{k}}_s.
\end{equation}

The subsequent section will describe a quantum high-pass filter that operates on the state vector obtained from this sequency-ordered Walsh-Hadamard transform. This filter will selectively suppress low-sequency components, allowing the output state to predominantly contain information about the edges present in the input grayscale image.

 \section{High-pass filter}\label{sec:high_pass_filter}

In signal processing, a high-pass filter is designed to pass components with frequencies (or sequencies in the Walsh domain) above a specific cutoff and attenuate components below it. For edge detection, this is important because high-sequency components in the Walsh domain encode information about sharp transitions and details, which correspond to edges in an image. Conversely, low-sequency components represent smooth, background features.

Consider a classical discrete signal $f = [f_0, f_1, f_2, \ldots, f_{N-1}]^T$, where $N = 2^n$ for some integer $n$. Let its sequency-ordered Walsh-Hadamard transformed counterpart be $\widehat{f} = [\widehat{f_0}, \widehat{f}_1, \widehat{f}_2, \ldots, \widehat{f}_{N-1}]^T$. In this sequency-ordered representation, $\widehat{f_0}$ corresponds to the least-sequency (DC) component, while $\widehat{f}_{N-1}$ represents the highest-sequency component. A classical high-pass filter selectively eliminates components below a chosen cutoff sequency $c$. The resulting filtered signal, $\widehat{f}_{HP}$, retains only the high-sequency components:
\begin{equation}
    \widehat{f}_{HP} = [0, 0, \ldots, 0, \widehat{f}_c, \widehat{f}_{c+1}, \ldots, \widehat{f}_{N-1}]^T.
\end{equation}

The quantum circuit for realizing this high-pass filter is adapted from the methodology presented in \cite{shukla2023quantum}. This quantum filter requires one ancilla qubit, which serves to tag or separate the low- and high-sequency components.

Let the initial quantum state of the system be $\ket{k_i} = \ket{0}_a \otimes \ket{k}_d$, where $\ket{0}_a$ is the ancilla qubit initialized to $\ket{0}$, and $\ket{k}_d$ represents the data qubits encoding the image information. From \mref{sec:state_preparation}, the data state $\ket{k}_d$ is the QPIE representation of the image:
\begin{equation*}
    \ket{k}_d = \frac{1}{\norm{f}}\sum_{m=0}^{N-1}f_m\ket{m},
\end{equation*}
where $\norm{f} = \sqrt{\sum_{m=0}^{N-1}f^{2}_m}$.

To prepare the state for filtering, we first apply the $X$ gate to the ancilla qubit and simultaneously apply the sequency-ordered Walsh-Hadamard transform $U_zH^{\otimes n}$ (discussed in \mref{sec:WHT}) to the data qubits. This composite operation transforms the initial state $\ket{k_i}$ into:
\begin{equation}
    (X_a \otimes U_zH^{\otimes n}_d) (\ket{0}_a \otimes \ket{k}_d) = \ket{1}_a \otimes (U_zH^{\otimes n}_d\ket{k}_d) = \ket{1}_a \otimes \ket{\widehat{k}}_s.
\end{equation}
Here, $\ket{\widehat{k}}_s = U_zH^{\otimes n}\ket{k}$ denotes the sequency-ordered Walsh-Hadamard transformed state of the image.

This state $\ket{1}_a \otimes \ket{\widehat{k}}_s$ is then fed into the quantum high-pass filter operator, $U_{HP}$. The action of $U_{HP}$ is defined such that it flips the ancilla qubit if the data qubits are in a low-sequency state ($m < c$), while leaving the ancilla unchanged for high-sequency states ($m \geq c$). Mathematically, this operation can be expressed as:
\begin{equation}
    U_{HP}(\ket{1}_a \otimes \ket{\widehat{k}}_s) = \left( (X_a \otimes \sum_{m<c}\ket{m}\bra{m}) + (I_a \otimes \sum_{m\geq c}\ket{m}\bra{m}) \right)(\ket{1}_a \otimes \ket{\widehat{k}}_s).
\end{equation}
The outcome of applying $U_{HP}$ is a superposition where the ancilla qubit tags the sequency components:
\begin{equation}
    U_{HP}(\ket{1}_a \otimes \ket{\widehat{k}}_s) = \ket{0}_a \otimes \ket{\widehat{k}}_{s,l} + \ket{1}_a \otimes \ket{\widehat{k}}_{s,h}.
\end{equation}
In this resulting state, $\ket{\widehat{k}}_{s,l}$ represents the low-sequency components of the transformed image (where the ancilla is now $\ket{0}_a$), and $\ket{\widehat{k}}_{s,h}$ represents the high-sequency components (where the ancilla remains $\ket{1}_a$). The operator $U_{HP}$ is implemented using a set of multi-controlled $X$-gates targeting the ancilla qubit, conditioned on the states of the data qubits. A more detailed discussion on the construction and operation of the high-pass filter operator $U_{HP}$ can be found in \cite{shukla2023quantum}.

\section{Proposed edge detection algorithm}\label{sec:edge_detection_method}

\begin{figure}[h!]
    \centering
    \includegraphics[width=\linewidth]{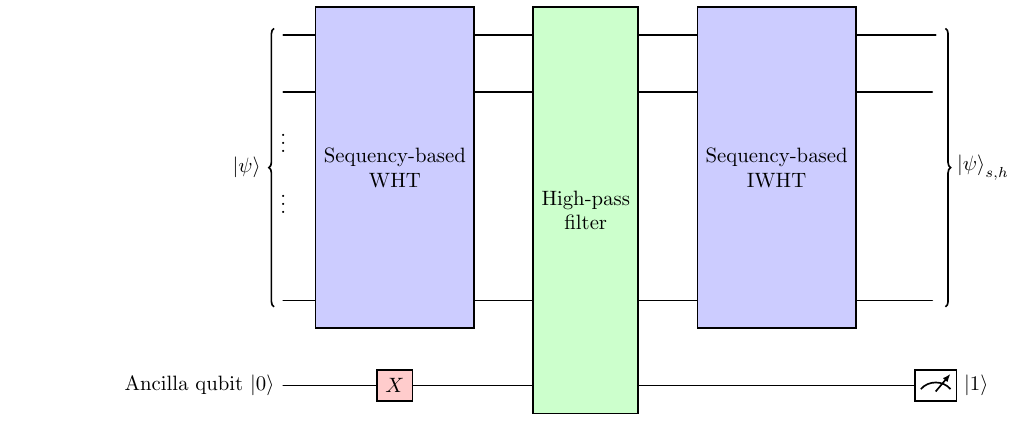}
    \caption{Block diagram for the proposed quantum edge detection approach.}\label{proposed_method_block_diagram}
\end{figure}

This section describes the complete quantum algorithm for edge detection, integrating the quantum tools and techniques discussed in previous sections. For an input image of size $N_1 \times N_2$, where $N_i=2^{n_i}$ for $n_i \in \mathbb{N}$, the algorithm requires a total of $n_1+n_2+1$ qubits. Of these, $n_1+n_2$ qubits are dedicated to encoding the image information using Quantum Probability Image Encoding (QPIE) (refer to \mref{sec:state_preparation}), and an additional ancilla qubit is used during the high-pass filtering process. A block diagram illustrating the proposed edge detection algorithm is presented in \mfig{proposed_method_block_diagram}.

\begin{algorithm}[H] \label{proposed_algorithm}
    \DontPrintSemicolon 
    \SetKwInput{KwInput}{Input} 
    \SetKwInput{KwOutput}{Output} 

    \KwInput{
    (a) A grayscale image of size $N_1 \times N_2$, where $N_i=2^{n_i}$ with $n_i \in \mathbb{N}$. \\
    (b) A chosen cutoff sequency $c$.
    }
    \KwOutput{An edge-detected image of size $N_1 \times N_2$.}

    \BlankLine 

    \KwSty{State Preparation:} \label{step:state_preparation}
    Prepare an initial quantum state $\ket{0}_a \otimes \ket{k}_d$, where the ancilla qubit $\ket{0}_a$ is initialized to $\ket{0}$, and the data qubits $\ket{k}_d$ encode the image information using Quantum Probability Image Encoding (QPIE). This step uses a total of $(n_1 + n_2 + 1)$ qubits.

    \BlankLine 

    \KwSty{Natural-Ordered Walsh-Hadamard Transform:} \label{step:natural_WHT}
    Apply an $X$ gate to the ancilla qubit $\ket{0}_a$ to flip it to $\ket{1}_a$, and simultaneously apply the Hadamard transform $H^{\otimes (n_1+n_2)}$ to the data qubits $\ket{k}_d$. The resulting state is $\ket{1}_a \otimes \ket{\widehat{k}}_d$, where $\ket{\widehat{k}}_d = H^{\otimes (n_1+n_2)}\ket{k}_d$ is the natural-ordered Walsh-Hadamard transform of the image state.

    \BlankLine 

    \KwSty{Sequency Ordering Transformation:} \label{step:sequency_transform}
    Apply the unitary operator $U_z$ to the data qubits. This operation transforms the natural-ordered Walsh-Hadamard state $\ket{\widehat{k}}_d$ into the desired sequency-ordered Walsh-Hadamard transformed state $\ket{\widehat{k}}_s$. The system state becomes $\ket{1}_a \otimes \ket{\widehat{k}}_s$.

    \BlankLine 

    \KwSty{High-Pass Filtering:} \label{step:high_pass_filter}
    Apply the quantum high-pass filter operator $U_{HP}$ to the system. This operator, composed of appropriate multi-controlled $X$ gates conditioned on the data qubits, separates the low and high-sequency components. The output state is $\ket{0}_a \otimes \ket{\widehat{k}}_{s,l} + \ket{1}_a \otimes \ket{\widehat{k}}_{s,h}$, where $\ket{\widehat{k}}_{s,l}$ represents the low-sequency components and $\ket{\widehat{k}}_{s,h}$ represents the high-sequency components in the sequency-ordered Walsh domain.

    \BlankLine 

    \KwSty{Inverse Sequency Ordering Transformation:} \label{step:inverse_sequency_transform}
    Apply the inverse unitary operator $U^{-1}_z$ to the data qubits. This transforms the sequency-ordered low-pass and high-pass components back to their natural-ordered Walsh-Hadamard representations: $\ket{0}_a \otimes \ket{\widehat{k}}_l + \ket{1}_a \otimes \ket{\widehat{k}}_{h}$.

    \BlankLine 

    \KwSty{Inverse Walsh-Hadamard Transform:} \label{step:inverse_WHT}
    Apply the inverse Hadamard transform $H^{\otimes (n_1+n_2)}$ to transform the quantum state back into the spatial (image) domain: $\ket{0}_a \otimes \ket{k}_l + \ket{1}_a \otimes \ket{k}_{h}$.

    \BlankLine 

     \KwSty{Measurement and Component Isolation:} \label{step:measurement}
    Perform a measurement on the ancilla qubit. By post-selecting on the ancilla qubit being in the state $\ket{1}_a$, the quantum state of the data qubits collapses to $\ket{k}_{h}$, which exclusively contains the high-sequency components, representing the edges of the image.

    \BlankLine 

    \KwSty{Image Reconstruction:} \label{step:reconstruction}
    Reconstruct the edge-detected image of size $N_1 \times N_2$ from the resulting quantum state using the inverse QPIE decoding technique.

    \BlankLine 
    \Return{The edge-detected image.}
    \caption{Algorithm for Edge Detection in Grayscale Images}
\end{algorithm}

\BlankLine 
We note that the proposed algorithm is applied to the original grayscale image to extract vertical edges and to its transpose to extract horizontal edges (ref.~Remark \ref{remark}). In \mref{subsec:computational_examples}, we will provide computational examples to illustrate edge detection using the proposed algorithm. 

\begin{remark} \label{remark}
    \leavevmode
    \begin{enumerate}[(i)]
        \item Algorithm~\ref{proposed_algorithm} processes image data to highlight features corresponding to vertical edges in the input matrix. To extract horizontal edge information, the image matrix is first transposed, and the algorithm is then applied to the transposed matrix. The resulting edge map is then transposed back to its original orientation, thereby detecting horizontal edges. The results from both applications is then is combined (e.g., by addition) to yield a comprehensive edge-detected image.
        Thus, constructing a complete edge map (capturing both vertical and horizontal edges) requires two executions of the algorithm, referred to as the vertical and horizontal passes. This approach is consistent with the methodology employed by the QHED algorithm, which also performs a transposition to detect horizontal edges.

        \item As part of the postprocessing steps, intensity scaling factors are applied to the output images from each pass to enhance the visual sharpness of the detected edges. Specifically, the pixel intensities of the output from Algorithm~\ref{proposed_algorithm} are scaled after both the vertical and horizontal passes.

        \item Unless mentioned otherwise, in all the computational examples presented in this work, a vertical scaling factor of~$3$ and a horizontal scaling factor of~$2$ were used. That is, the intensity matrix obtained after the vertical pass of Algorithm~\ref{proposed_algorithm} is multiplied by~$3$, and the output from the horizontal pass is multiplied by~$2$.
    \end{enumerate}
\end{remark}

\subsubsection{Quantum circuits}

\begin{figure}[H]
    \centering
    \begin{subfigure}[b]{\textwidth}
        \centering
        \includegraphics[scale=0.25]{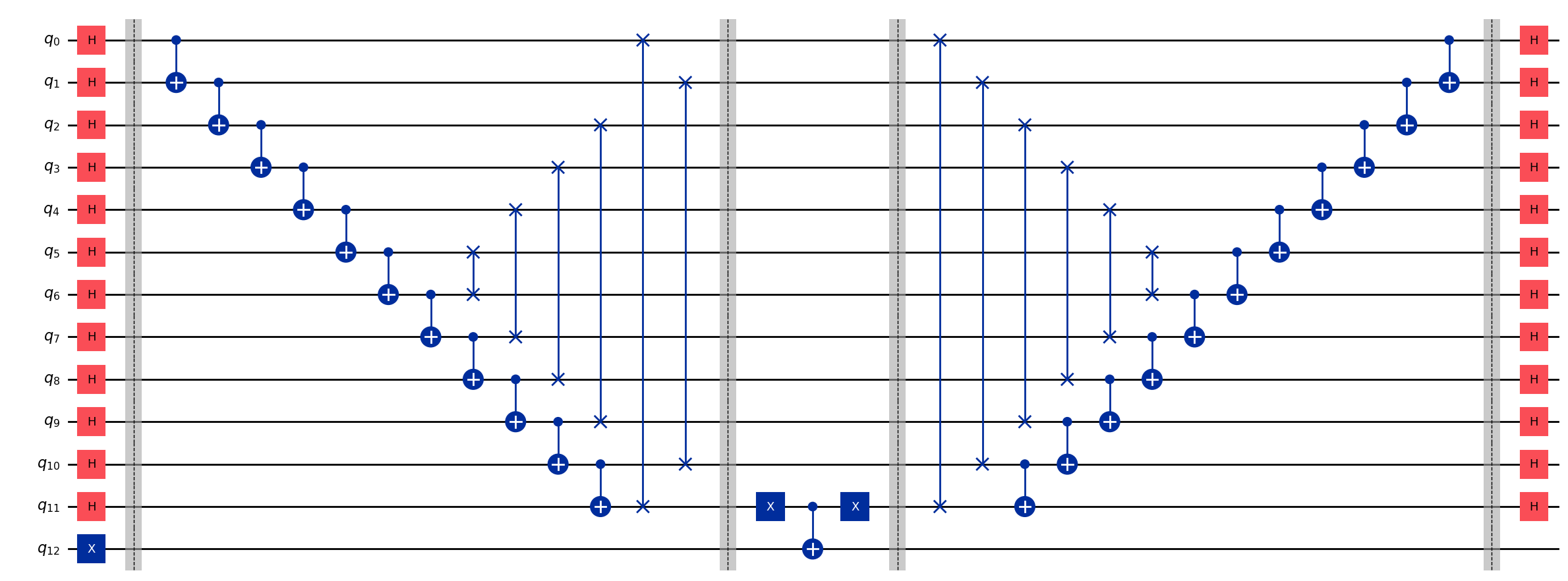}
        \caption{Quantum circuit for edge detection on a $64 \times 64$ image using the proposed algorithm with cutoff sequency $\frac{N}{2}$.}
        \label{fig:quantum_circuit_Nby2}
    \end{subfigure}
    \vfill
    \begin{subfigure}[b]{\textwidth}
        \centering
        \includegraphics[scale=0.25]{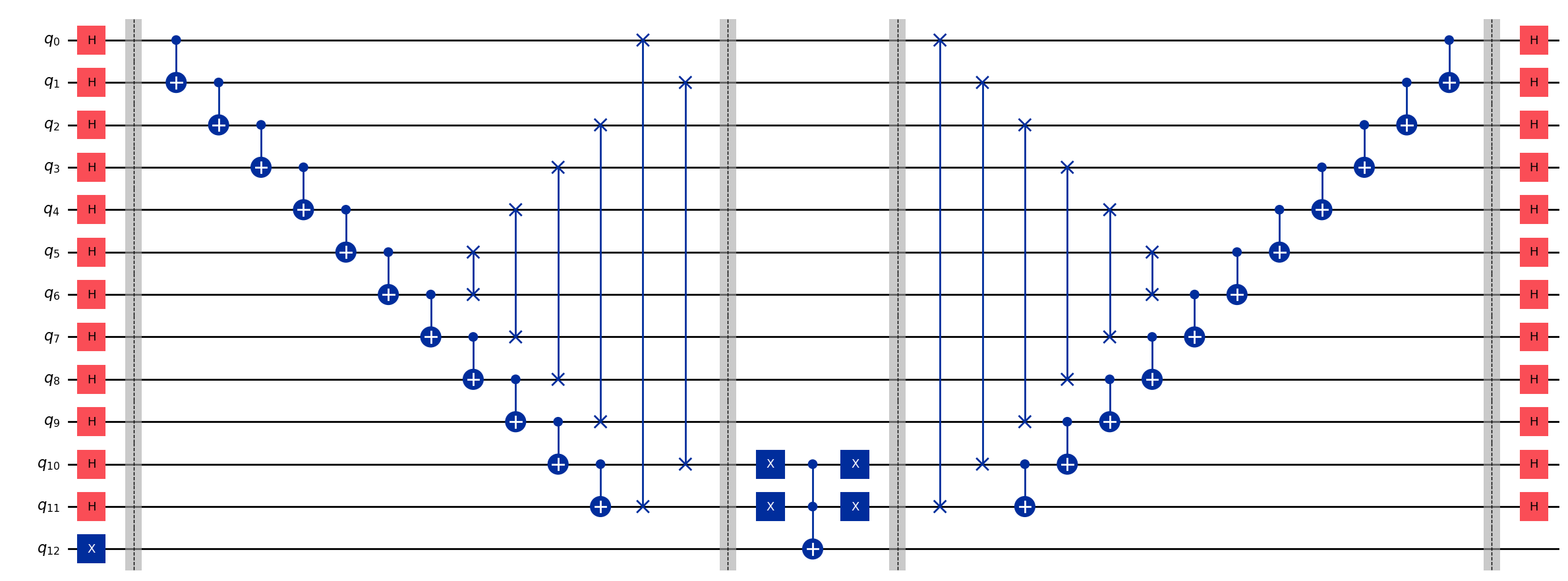}
        \caption{Quantum circuit for edge detection on a $64 \times 64$ image using the proposed algorithm with cutoff sequency $\frac{N}{4}$.}
        \label{fig:quantum_circuit_Nby4}
    \end{subfigure}
    \caption{Quantum circuits for edge detection using Algorithm~\ref{proposed_algorithm}. Each circuit consists of three main stages: (i) the sequency-ordered Walsh–Hadamard transform (WHT), (ii) high-pass filtering, and (iii) the inverse WHT. The filtering stage (between the second and third barriers) is parameterized by the cutoff sequency.}
    \label{fig:quantum_circuits}
\end{figure}

The quantum circuits implementing Algorithm~\ref{proposed_algorithm} for edge detection on a $64 \times 64$ grayscale image are shown in Figure~\ref{fig:quantum_circuits}. In this example, a total of $n = 6 + 6 = 12$ qubits are used to represent the flattened image, along with one additional ancilla qubit required for the high-pass filtering operation within Algorithm~\ref{proposed_algorithm}. 

Each circuit consists of three main stages: the sequency-ordered Walsh–Hadamard transform (WHT), the high-pass filtering stage, and the inverse WHT. The segment before the second barrier corresponds to the WHT, while the segment after the third barrier implements its inverse. The high-pass filtering operation, which corresponds to Step~4 of Algorithm~\ref{proposed_algorithm}, is applied between the second and third barriers.

Figure~\ref{fig:quantum_circuit_Nby2} shows the circuit configuration for a cutoff sequency of $\frac{N}{2}$, whereas Figure~\ref{fig:quantum_circuit_Nby4} shows the corresponding circuit for a lower cutoff sequency of $\frac{N}{4}$. The circuit structure remains the same in both cases, with only the filtering parameters modified to achieve the desired cutoff threshold.

\subsection{Computational examples}\label{subsec:computational_examples}

This section presents computational examples to validate the proposed edge detection Algorithm \ref{proposed_algorithm} and compare its performance against the Quantum Hadamard Edge Detection (QHED) algorithm, as described in \cite{yao2017quantum}. For quantitative comparison, we use the Structural Similarity Index Measure (SSIM), which evaluates the perceptual similarity between two images by considering structural correlations, luminosity, and contrast \cite{shukla2022hybrid}.

\begin{figure}[H]
    \centering
    \begin{subfigure}[t]{.45\textwidth}
        \centering
        \includegraphics[width=\linewidth]{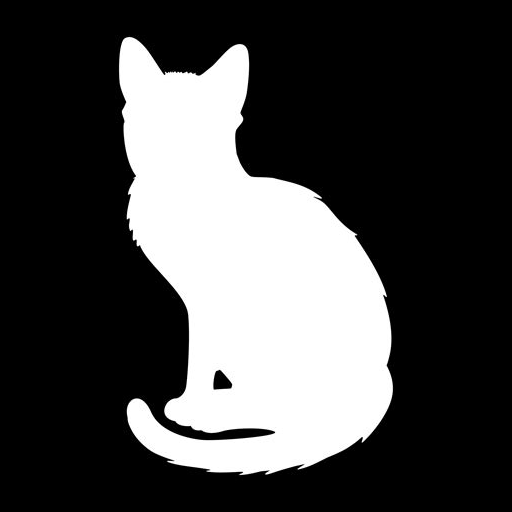}
        \caption{Original $512\times 512$ grayscale image of a cat silhouette.}\label{img:original_cat}
    \end{subfigure}\hfill
    \begin{subfigure}[t]{.45\textwidth}
        \centering
        \includegraphics[width=\linewidth]{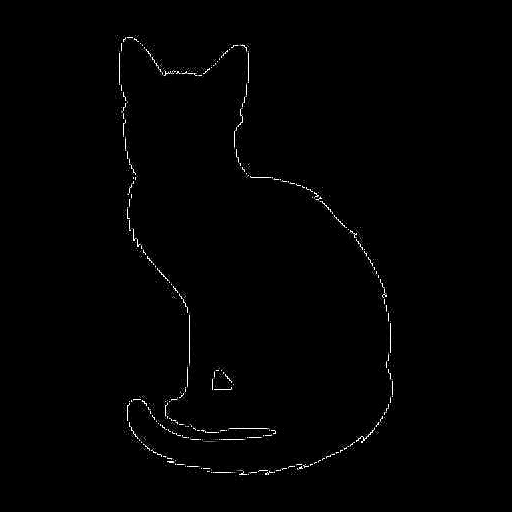}
       \caption{Detected edges using the proposed algorithm with cutoff sequency $\frac{N}{2}$. A vertical scaling factor of~$9$ and a horizontal scaling factor of~$6$ were used to enhance the pixel intensity of the detected edges.}
\label{img:cat_edge_detected}
    \end{subfigure}
   \caption{Edge detection results on a $512 \times 512$ grayscale image of a cat silhouette using Algorithm~\ref{proposed_algorithm}. Both vertical and horizontal passes of the algorithm were applied, and the resulting edge maps were combined to form the final edge-detected image.}
\label{img:cat_edge_detection}
\end{figure}

Figure~\ref{img:cat_edge_detection} illustrates the application of Algorithm~\ref{proposed_algorithm} on a $512 \times 512$ grayscale image of a cat silhouette. The original input image is shown in Figure~\ref{img:original_cat}, while the corresponding edge-detected result is presented in Figure~\ref{img:cat_edge_detected}. 
The edge detection process was performed using both the vertical and horizontal passes of the algorithm, as outlined in Remark~\ref{remark}. Specifically, the input image was processed in its original orientation to extract vertical edges, followed by a transposed version of the image to extract horizontal edges. The resulting edge maps from the two passes were subsequently combined to form the final edge-detected image. 
As evident from Figure~\ref{img:cat_edge_detected}, the proposed algorithm successfully highlights the prominent contours of the cat silhouette, demonstrating its effectiveness in identifying clear edge structures within the image. The cutoff sequency parameter was set to $\frac{N}{2}$ for this example, which yielded visually sharp and distinct edge features. 

\begin{figure}[H]
	\centering
    \begin{subfigure}[t]{.3\textwidth}
        \centering
        \includegraphics[width=\linewidth]{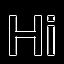}
        \caption{Original $64\times 64$ computer-generated string image.}\label{img_string}
    \end{subfigure}\hfill
    \begin{subfigure}[t]{.3\textwidth}
        \centering
        \includegraphics[width=\linewidth]{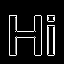}
        \caption{Edge detection using proposed algorithm with cutoff sequency $\frac{N}{2}$.}\label{img_string_WHT_N2}
    \end{subfigure}\hfill
    \begin{subfigure}[t]{.3\textwidth}
        \centering
        \includegraphics[width=\linewidth]{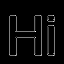}
        \caption{Edge detection using QHED.}\label{img_string_QHED}
    \end{subfigure}\\[0.5cm] 

    \begin{subfigure}[t]{.3\textwidth}
        \centering
        \includegraphics[width=\linewidth]{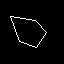}
        \caption{Original $64\times 64$ computer-generated polygon image.}\label{img_polygon}
    \end{subfigure}\hfill
    \begin{subfigure}[t]{.3\textwidth}
        \centering
        \includegraphics[width=\linewidth]{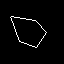}
        \caption{Edge detection using proposed algorithm with cutoff sequency $\frac{N}{2}$.}\label{img_polygon_WHT_N2}
    \end{subfigure}\hfill
    \begin{subfigure}[t]{.3\textwidth}
        \centering
        \includegraphics[width=\linewidth]{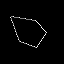}
        \caption{Edge detection using QHED.}\label{img_polygon_QHED}
    \end{subfigure}\\[0.5cm] 

    \begin{subfigure}[t]{.3\textwidth}
        \centering
        \includegraphics[width=\linewidth]{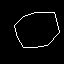}
        \caption{Original $64\times 64$ computer-generated polygon image.}\label{img_polygon2}
    \end{subfigure}\hfill
    \begin{subfigure}[t]{.3\textwidth}
        \centering
        \includegraphics[width=\linewidth]{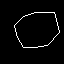}
        \caption{Edge detection using proposed algorithm with cutoff sequency $\frac{N}{2}$.}\label{img_polygon2_WHT_N4}
    \end{subfigure}\hfill
    \begin{subfigure}[t]{.3\textwidth}
        \centering
        \includegraphics[width=\linewidth]{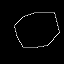}
        \caption{Edge detection using QHED.}\label{img_polygon2_QHED}
    \end{subfigure}

    \caption{Edge detection comparison on $64\times 64$ computer-generated grayscale images. 
}\label{img:comparison_images}
\end{figure}

The next set of computational examples involves computer-generated grayscale images, which serve as ground truth images for SSIM evaluation. \mfig{img:comparison_images} illustrates the visual comparison between the proposed algorithm and QHED for various computer-generated patterns, including strings and polygons.

Table~\ref{table:comparison} presents a comparative evaluation of the proposed algorithm (Algorithm~\ref{proposed_algorithm}) and the QHED approach for edge detection on various computer-generated images. For each example, the detected edge maps obtained from both methods are compared against the original image using the Structural Similarity Index Measure (SSIM). 

As seen from the table, the proposed algorithm consistently achieves similar or higher SSIM scores across all test cases. For instance, in the case of the string-shaped image (Figure~\ref{img_string}), the proposed algorithm yields an SSIM of $0.3791$ compared to $0.3654$ for QHED. Similar improvements are observed for the polygon-shaped images shown in Figures~\ref{img_polygon} and~\ref{img_polygon2}, where the proposed method achieves SSIM scores of $0.7255$ and $0.6794$, respectively, both slightly exceeding the corresponding QHED results.

In addition to the quantitative improvements, the edge maps produced by the proposed algorithm exhibit slightly sharper visual features and better preserve the structural characteristics of the original images, as illustrated in the corresponding figures. These results demonstrate that the proposed approach achieves comparable or slightly superior edge detection performance relative to the QHED method, both in terms of structural similarity and visual quality. The primary advantage of the proposed approach, as outlined in Algorithm~\ref{proposed_algorithm}, lies in its significantly lower computational cost compared to the QHED method, while delivering similar or marginally better edge detection results (ref.~\mref{sec:computational_cost}).

\begin{table}[h]
    \centering
    \renewcommand{\arraystretch}{1.5}
    \begin{tabular}{|c|c|c|c|}
        \hline
        \textbf{Original Image} & \textbf{Detected Edges (Output Image)} & \textbf{Algorithm} & \textbf{SSIM} \\
        \hline
        Figure \ref{img_string} & Figure \ref{img_string_WHT_N2} & Algorithm \ref{proposed_algorithm} & 0.3791\\
        \hline
        Figure \ref{img_string} & Figure \ref{img_string_QHED} & QHED & 0.3654\\
        \hline
        Figure \ref{img_polygon} & Figure \ref{img_polygon_WHT_N2} & Algorithm \ref{proposed_algorithm} & 0.7255\\
        \hline
        Figure \ref{img_polygon} & Figure \ref{img_polygon_QHED} & QHED & 0.7228 \\
        \hline
        Figure \ref{img_polygon2} & Figure \ref{img_polygon2_WHT_N4} & Algorithm \ref{proposed_algorithm} & 0.6794\\
        \hline
        Figure \ref{img_polygon2} & Figure \ref{img_polygon2_QHED} & QHED & 0.6715\\
        \hline
    \end{tabular}
    \caption{Comparison of SSIM scores for edge detection on test images using our proposed algorithm and QHED algorithm.}\label{table:comparison}
\end{table}

\begin{figure}[H]
    \centering
    \begin{subfigure}[b]{0.33\textwidth}
        \centering
        \includegraphics[scale=0.4]{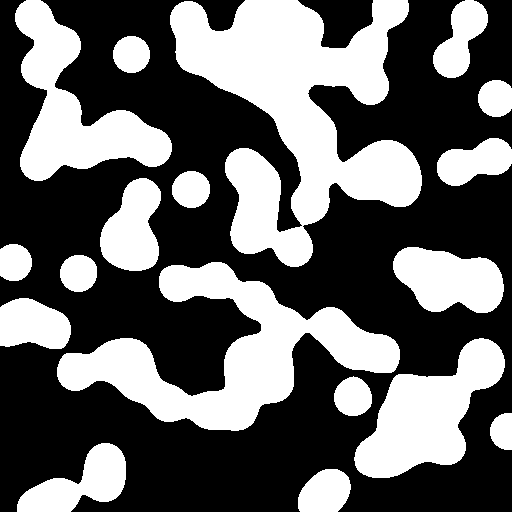}
        \caption{Original $512 \times 512$ binary blobs image.}
        \label{fig:binary_blob_original}
    \end{subfigure}
    \hfill
    \begin{subfigure}[b]{0.33\textwidth}
        \centering
        \includegraphics[scale=0.4]{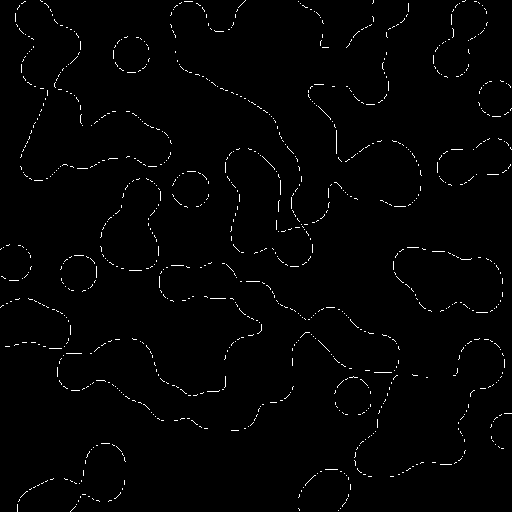}
        \caption{Detected edges using the proposed algorithm with cutoff sequency $\frac{N}{2}$.}
        \label{fig:binary_blob_N_2}
    \end{subfigure}
    \hfill
    \begin{subfigure}[b]{0.33\textwidth}
        \centering
        \includegraphics[scale=0.4]{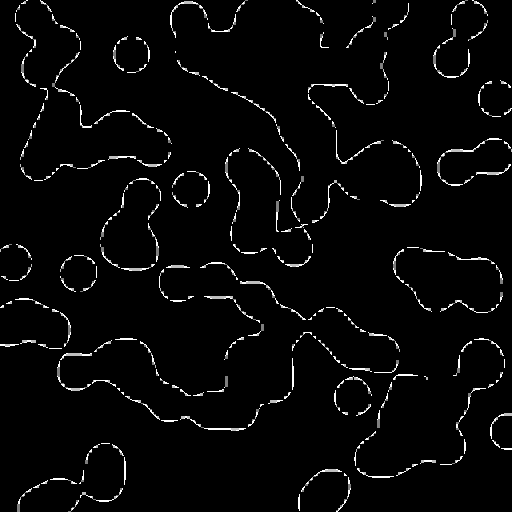}
        \caption{Detected edges using the proposed algorithm with cutoff sequency $\frac{N}{4}$.}
        \label{fig:binary_blob_N_4}
    \end{subfigure}
    \vfill
    \begin{subfigure}[b]{0.33\textwidth}
        \centering
        \includegraphics[scale=0.4]{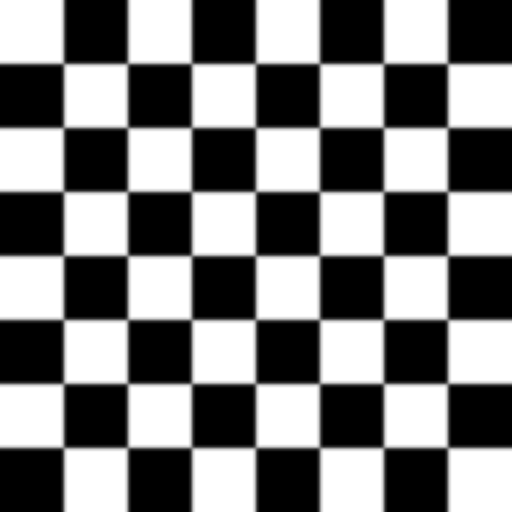}
        \caption{Original $512 \times 512$ checkerboard image.}
        \label{fig:checkerboard_original}
    \end{subfigure}
    \hfill
    \begin{subfigure}[b]{0.33\textwidth}
        \centering
        \includegraphics[scale=0.4]{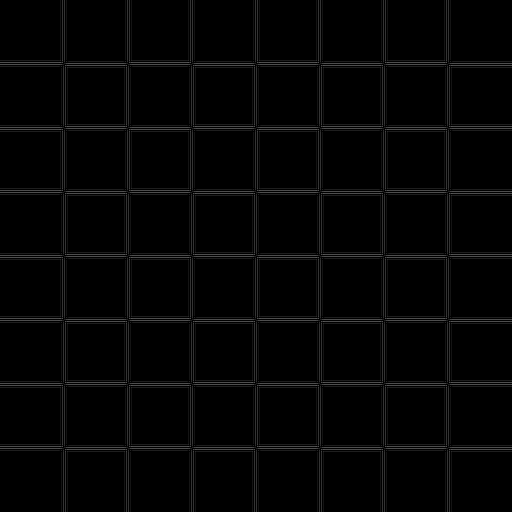}
        \caption{Detected edges using the proposed algorithm with cutoff sequency $\frac{N}{2}$.}
        \label{fig:checkerboard_N_2}
    \end{subfigure}
    \hfill
    \begin{subfigure}[b]{0.33\textwidth}
        \centering
        \includegraphics[scale=0.4]{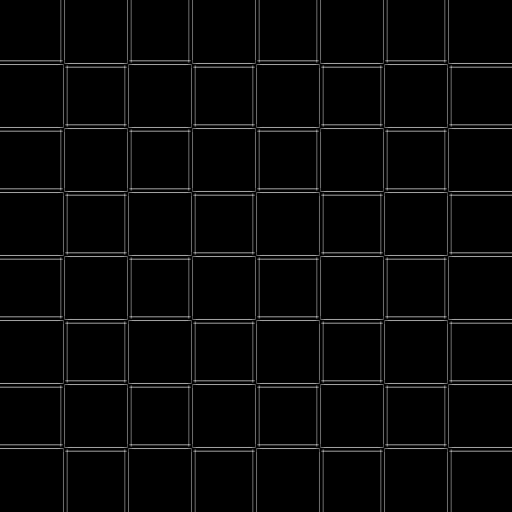}
        \caption{Detected edges using the proposed algorithm with cutoff sequency $\frac{N}{4}$.}
        \label{fig:checkerboard_N_4}
    \end{subfigure}
    
    \caption{Edge detection results using Algorithm~\ref{proposed_algorithm} on two test images: binary blobs (top row) and a checkerboard pattern (bottom row). For each image, results are shown for two cutoff sequency values, $\frac{N}{2}$ and $\frac{N}{4}$.}
    \label{fig:examples}
\end{figure}

The effectiveness of Algorithm~\ref{proposed_algorithm} in detecting edges is further illustrated in Figure~\ref{fig:examples} using two test images, a binary blobs image and a checkerboard pattern. Figures~\ref{fig:binary_blob_original} and~\ref{fig:checkerboard_original} show the original $512 \times 512$ input images. The corresponding edge-detected outputs for cutoff sequency values of $\frac{N}{2}$ and $\frac{N}{4}$ are shown in Figures~\ref{fig:binary_blob_N_2}--\ref{fig:binary_blob_N_4} and~\ref{fig:checkerboard_N_2}--\ref{fig:checkerboard_N_4}, respectively. As the cutoff sequency decreases from $\frac{N}{2}$ to $\frac{N}{4}$, the resulting edge maps capture finer image details, highlighting the tunable nature of the proposed approach.

\subsection{Computational complexity}
\label{sec:computational_cost}

For edge detection in a grayscale image of size $N_1 \times N_2$, our proposed algorithm achieves a computational complexity of $\mathcal{O}(\log_2(N_1N_2))$, representing an improvement over the QHED approach, which has a complexity of $\mathcal{O}(\mathrm{poly}(\log_2(N_1N_2)))$.

We note that the quantum image representation used by QHED \cite{yao2017quantum} is Quantum Probability Image Encoding (QPIE), which is also employed by the proposed algorithm. The core operation in QHED for achieving edge detection is quantum amplitude permutation. To briefly explain this, we consider a $(n+1)$-qubit state $\ket{f} = \frac{1}{\sqrt{2}}(c_0,~c_0,~c_1,~c_1,~\ldots,~c_{N-1},~c_{N-1})^T$, where $N=2^n$ and $c_i$ are the pixel intensity values. Let $D_{2^{n+1}}$ be the operator for quantum amplitude permutation. Its action on the state $\ket{f}$ is:
\begin{equation}
D_{2^{n+1}}\ket{f} = \frac{1}{\sqrt{2}}(c_0,~c_1,~c_1,~\ldots,~c_{N-1},~c_{N-1},~c_0)^T.
\end{equation}
In matrix form, the quantum amplitude permutation operator $D_{2^{n+1}}$ is given by a $2^{n+1} \times 2^{n+1}$ matrix:
\begin{equation*}
D_{2^{n+1}} =
	\begin{pmatrix*}[r]
		0 & 1 & 0 & \cdots & 0 & 0  \\
		0 & 0 & 1 & \cdots & 0 & 0  \\
		0 & 0 & 0 & \cdots & 0 & 0  \\
		\vdots & \vdots & \vdots & \ddots & \vdots & \vdots  \\
		0 & 0 & 0 &  & 0 & 1  \\
		1 & 0 & 0 & \cdots & 0 & 0  \\
	\end{pmatrix*}.
\end{equation*}

The computational cost to efficiently perform quantum amplitude permutation is $\mathcal{O}(\text{poly}(n))$, where $n$ is the number of qubits encoding the image. In contrast, the most computationally significant step in our proposed algorithm is the sequency-ordered Walsh-Hadamard transform, which has a computational cost of $\mathcal{O}(n)$. Consequently, for a grayscale image of size $N_1 \times N_2$, the proposed algorithm has a computational cost of $\mathcal{O}(\text{log}_2(N_1N_2))$, which is an improvement over QHED's $\mathcal{O}(\text{poly}(\text{log}_2(N_1N_2)))$. It is important to note that in both algorithms, the costs associated with quantum state preparation using QPIE and the partial measurement of the ancilla qubit are not included in these complexity analyses.

For many quantum algorithms, including Algorithm \ref{proposed_algorithm} presented in this work, state preparation and measurement remain significant bottlenecks in achieving a practical quantum advantage over classical methods. While there are special cases where certain states can be prepared efficiently \cite{shukla2024efficient}, in general, preparing the quantum representation of an image of size $N_1 \times N_2$ requires $\mathcal{O}(N_1 N_2)$ operations, eliminating any computational advantage over classical approaches. However, the proposed algorithm becomes advantageous when the input state is the output of a preceding quantum subroutine. Furthermore, one can extract the global features of the filtered image without the overhead of fully measuring the quantum state. For instance, one can apply the swap test to compare the detected edges directly with a known template, avoiding the need for complete reconstruction of the filtered image.

\section{Conclusion}\label{sec:conclusion}

This work introduces a novel quantum algorithm for edge detection in digital grayscale images. A key advantage of the proposed algorithm lies in its improved computational complexity, as it achieves a quantum circuit depth of $\mathcal{O}(\text{log}_2(N_1N_2))$ for an $N_1\times N_2$ image, compared to the Quantum Hadamard Edge Detection (QHED) algorithm which has a depth of $\mathcal{O}(\text{poly}(\text{log}_2(N_1N_2)))$. Both the proposed algorithm and QHED exclude the cost of quantum state preparation using Quantum Probability Image Encoding (QPIE) and the partial measurement of the ancilla qubit in their complexity analyses. The primary source of this computational advantage stems from the efficient implementation of the sequency-ordered Walsh-Hadamard transform, which requires a quantum circuit depth and gate-complexity of $\mathcal{O}(n)$ for an $n$-qubit system.

The efficacy of the proposed algorithm has been validated through computational examples. The results from these examples confirm the algorithm's capability in accurately detecting edges and highlight its improved efficiency.

\FloatBarrier

\bibliographystyle{unsrt}

\end{document}